\documentclass[pdflatex,sn-nature]{sn-jnl}

\usepackage{graphicx}%
\usepackage{multirow}%
\usepackage{amsmath,amssymb,amsfonts}%
\usepackage{amsthm}%
\usepackage{mathrsfs}%
\usepackage[title]{appendix}%
\usepackage{xcolor}%
\usepackage{textcomp}%
\usepackage{manyfoot}%
\usepackage{booktabs}%
\usepackage{algorithm}%
\usepackage{algorithmicx}%
\usepackage{algpseudocode}%
\usepackage{listings}%
	
\theoremstyle{thmstyleone}%
\theoremstyle{thmstyletwo}%

\theoremstyle{thmstylethree}%

\begin{document}

\title[High Harmonic Generation by Bright Squeezed Vacuum]{High Harmonic Generation by Bright Squeezed Vacuum}


\author[1,2]{\fnm{Andrei} \sur{Rasputnyi}}

\author[3,4,5]{\fnm{Zhaopin} \sur{Chen}}

\author[3,4,5]{\fnm{Michael} \sur{Birk}}

\author[3,4,5]{\fnm{Oren} \sur{Cohen}}

\author[4,5,6]{\fnm{Ido} \sur{Kaminer}}

\author[3,4,5]{\fnm{Michael} \sur{Kr\"uger}}

\author[7]{\fnm{Denis} \sur{Seletskiy}}

\author*[1,2,6]{\fnm{Maria} \sur{Chekhova}}\email{maria.chekhova@mpl.mpg.de}

\author[1,8]{\fnm{Francesco} \sur{Tani}}

\affil[1]{\orgdiv{Max Planck Institute for the Science of Light}, \orgaddress{\street{Staudtstr. 2}, \city{Erlangen}, \postcode{91058}, \country{Germany}}}

\affil[2]{\orgdiv{Friedrich-Alexander Universität Erlangen-Nürnberg}, \orgaddress{\street{Staudtstr. 7/B2}, \city{Erlangen}, \postcode{91058}, \country{Germany}}}

\affil[3]{\orgdiv{Department of Physics, Technion -- Israel Institute of Technology}, \orgaddress{\street{Technion City}, \city{Haifa}, \postcode{32000}, \country{Israel}}}

\affil[4]{\orgdiv{Solid State Institute, Technion -- Israel Institute of Technology}, \orgaddress{\street{Technion City}, \city{Haifa}, \postcode{32000}, \country{Israel}}}

\affil[5]{\orgdiv{Helen Diller Quantum Center, Technion -- Israel Institute of Technology}, \orgaddress{\street{Technion City}, \city{Haifa}, \postcode{32000}, \country{Israel}}}

\affil[6]{\orgdiv{Faculty of Electrical and Computer Engineering,
Technion -- Israel Institute of
Technology}, \orgaddress{\street{Technion City}, \city{Haifa}, \postcode{32000}, \country{Israel}}}

\affil[7]{\orgdiv{femtoQ Laboratory, Department of Engineering Physics, Polytechnique Montréal}, \orgaddress{\city{Montréal}, \postcode{Québec H3T 1J4}, \country{Canada}}}

\affil[8]{\orgdiv{Univ. Lille, CNRS, UMR 8523—PhLAM—Physique des Lasers Atomes et Molécules}, \orgaddress{\city{Lille}, \postcode{F-59000}, \country{France}}}


\abstract{We observe non-perturbative high harmonic generation in solids driven by a macroscopic quantum state of light, bright squeezed vacuum (BSV), which we generate in a single spatiotemporal mode. The BSV-driven process is considerably more efficient in the generation of high harmonics than classical light of the same mean intensity. Due to its broad photon-number distribution, covering states from $0$ to $2 \times 10^{13}$ photons per pulse, and sub-cycle electric field fluctuations over $\pm1\hbox{V}/\hbox{\r{A}}$, BSV provides access to free carrier dynamics within a much broader range of peak intensities than accessible with classical light. 
Our findings contribute to recent developments of quantum optics with extreme intensities, moving beyond its traditional focus on low photon numbers, and providing a new method for exploring extreme nonlinearities in solids.}




\maketitle

\section{Bright squeezed vacuum as a driver of high harmonic generation}\label{sec1}

High harmonic generation (HHG) lies at the foundation of attosecond science, extreme nonlinear optics, and an increasing number of applications \cite{Ferray1988,McPherson1987,Krausz2009}. Initially observed by irradiating gases \cite{McPherson1987, Ferray1988,Krausz2009, Popmintchev2012}, with intense ultrashort near-infrared (IR) pulses, nowadays, harmonics are generated also from liquids~\cite{Luu2018, Mondal23} and solids~\cite{Ghimire2011,Vampa2015, Goulielmakis2022}, using pump lasers with wavelengths spanning from the far-IR to the ultraviolet (UV) \cite{Ganeev2007, Popmintchev2015, Marceau_2017}. 
This non-perturbative process has applications for sources delivering high-photon energy radiation reaching beyond the x-ray water window~\cite{Popmintchev2012, Teichmann2016, Cousin2014}, and for creating attosecond light pulses~\cite{Paul2001,Hentschel2001}. By leveraging on its intrinsic sub-cycle nature, HHG is now being exploited for observing charge dynamics in matter at unprecedentedly short timescales, providing radically new and powerful tools for studying phenomena such as multi-electron correlations, phase transitions, and topological effects~\cite{Imai2020, Schmid2021, Heide2022, Shao2022}. As a result, over the past decade, HHG in solids has been used for a wide range of studies such as all-optical energy band structure retrieval \cite{Vampa2015, Hohenleutner2015}, extreme ultraviolet spectroscopy \cite{Tzallas2011, Luu2015, Nisoli2017}, laser picoscopy of the valence electron structure~\cite{Lakhotia2020}, investigations of quantum phase transitions~\cite{Alcala2022}, and Berry phases~\cite{Uzan2024}.

Despite the wide range of experimental conditions under which it has been observed and its many applications, HHG has been described mainly in semiclassical terms. A precise description of HHG depends on the quantum properties of matter (e.g., solid crystals or gas molecules). However, the intense light field driving HHG is usually considered to be classical.
Meanwhile, recent works have revealed quantum-optical features of the high harmonics~\cite{Gorlach2020,sloan2023entangling, yi2024generation,Stammer2024} and of their driving field following their generation~\cite{Tsatrafyllis2017, Tsatrafyllis2019, Lewenstein2021}. Still, all these works deal with the HHG from classical (coherent) pulses of light. The use of quantum light for driving HHG has been considered theoretically~\cite{Gorlach2023, EvenTzur2023} but not realized, since HHG requires intensities on the order of $1\hbox{ TW}\ \hbox{cm}^{-2}$, or photon numbers about $10^{13}$, in a short pulse of just tens of femtoseconds or shorter. Such extreme conditions have remained so far inaccessible for quantum state engineering. Indeed, quantum states of light produced in quantum optics typically contain only a few photons per mode. Macroscopic quantum states can be obtained by adding a strong coherent component to a squeezed vacuum \cite{Vahlbruch2016} or to a single photon \cite{Bruno2013, Lvovsky2013}, but even in this case, most of the energy is contained in the classical part \cite{LAGHAOUT201596}.

\begin{figure}[h]
\centering
\includegraphics[width=0.8\textwidth]{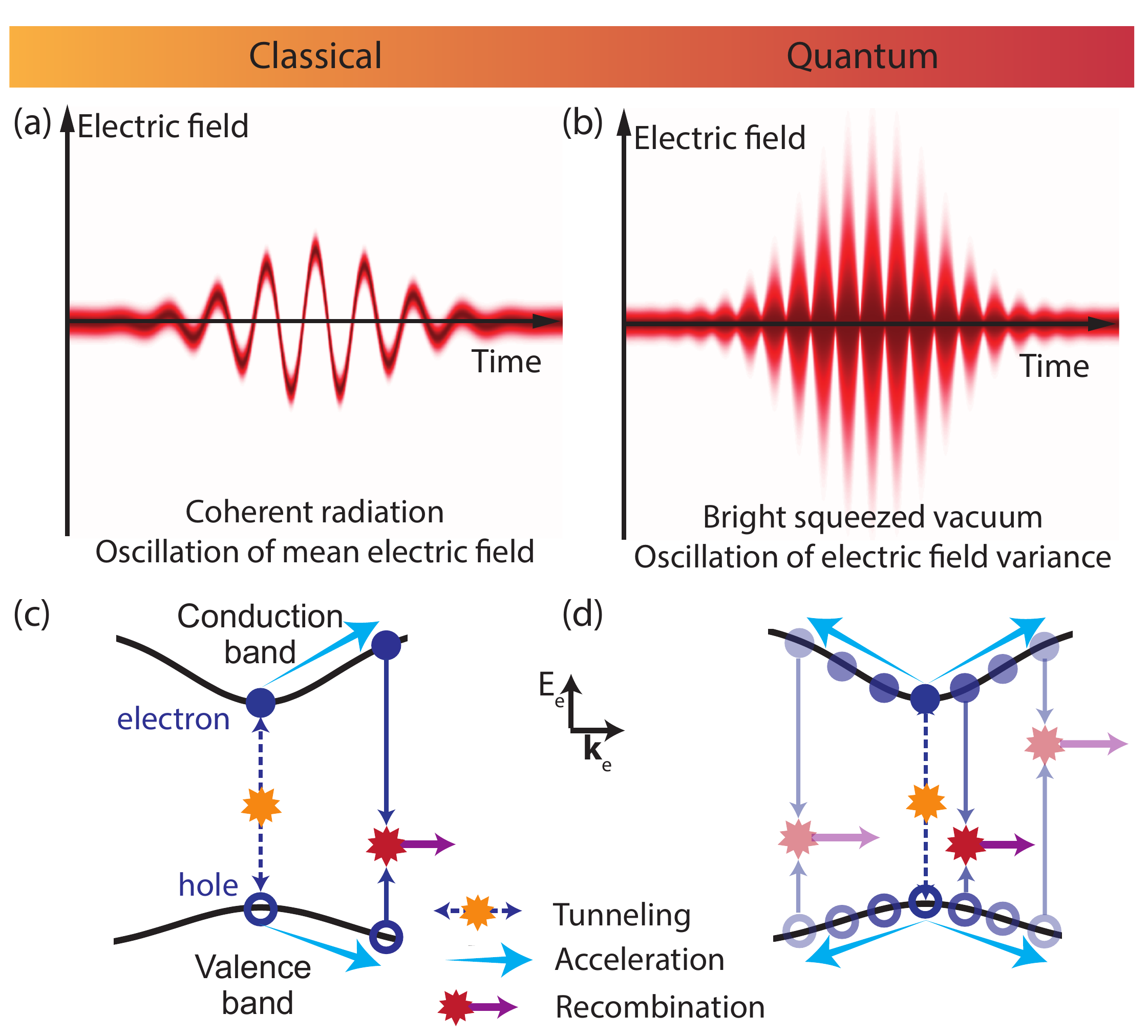}
\caption{(a) The electric field of a coherent pulse exhibits a mean value oscillating at the optical frequency and a constant variance. (b) The electric field of a bright squeezed vacuum pulse: the  mean value is constantly zero, while the variance oscillates at twice the carrier frequency. (c) High harmonic generation by coherent light in a semiconductor in the interband mechanism. An electron-hole pair emerges via perturbative (multiphoton) or non-perturbative excitation, is accelerated by the electric field, gaining energy, and finally recombines, emitting high-energy photons. (d) High harmonic generation by bright squeezed vacuum in a semiconductor. Due to the extremely large uncertainty of the BSV field during half a cycle, the high harmonic generation is driven by a continuum of paths for electron-hole pairs.}\label{fig1}
\end{figure}

An exception is bright squeezed vacuum (BSV), a macroscopic quantum state of light generated at the output of a strongly pumped unseeded optical parametric amplifier. 
BSV is a quantum superposition of even-photon-number states extending to high photon numbers. Recent experiments have achieved BSV pulses with mean photon number $\langle N\rangle$ as high as $10^{13}$~\cite{Spasibko2017, Manceau2019},  and with the photon-number probability distribution so broad that photon numbers exceeding the mean by a factor of 6 or more occurred with 2 \% probability.   
This feature is in striking contrast with the photon-number distribution of classical (coherent-state) light, whose width is as small as $\Delta N=\sqrt{\langle N\rangle}$ (the shot-noise limit). This implies that for classical light with $\langle N\rangle=10^{13}$, the probability of an event exceeding the mean by a factor of 6 or more is $\sim 10^{-10^{13}}$.

In addition, BSV has been shown to manifest non-classical features as sub-shot-noise photon-number correlations \cite{Finger2015} and polarization entanglement \cite{Iskhakov2012a}. It was also predicted to enhance sub-cycle sensing of quantum optical fields \cite{Virally2021}.
The broad photon-number probability distribution of BSV leads to superbunching \cite{Iskhakov2012}: its intensity correlation function of order $n$, $g^{(n)}\equiv \langle:N^n:\rangle/\langle N\rangle^n$, is considerably higher than for thermal light. This feature becomes more pronounced at large $n$: while for thermal light, the correlation functions are $g_{th}^{(n)}=n!$, for BSV they are $g_{BSV}^{(n)}=(2n-1)!!$. Superbunching enhances multiphoton effects, such as perturbative harmonic generation~\cite{Spasibko2017} or nonlinear electron emission from nanotips \cite{heimerl2023multiphoton}.

All these properties suggest that BSV is an intriguing and interesting alternative driver for the harmonics providing a new route to combine strong-field physics and quantum optics, exploring new phenomena accessible only with quantum light, and, possibly, enhancing HHG. Recently, these prospects have triggered theoretical works proposing applications of BSV in strong-field physics, predicting an extended plateau for HHG \cite{Gorlach2023}, modification of electron dynamics \cite{EvenTzur2023, EvenTzur2024}, HHG selection rules~\cite{Hegazi2023}, and quadrature squeezing in the extreme ultraviolet spectral range \cite{tzur2023generation}. 

In this work, we report the first observation of HHG driven by quantum light. Specifically, we demonstrate BSV-driven high harmonics from solids and show that compared to classical light, BSV enhances the harmonic yield, revealing a boost of multiphoton processes. These experimental results are supported by numerical investigations, that extend BSV-driven HHG simulations in gases \cite{Gorlach2020, Gorlach2023} to solids and show that these can reproduce the main trend. Furthermore, we show that the sparse nature of BSV light provides a simple and convenient route for suppressing sample damage, enabling to probe electron dynamics in solids in a previously inaccessible regime. 

In contrast to a bright classical light pulse, which has a well-defined amplitude of the electric field with very weak fluctuations (Fig.~\ref{fig1}(a)), the mean electric field of the BSV remains at zero for all times, while its variance oscillates at twice the light frequency (Figure \ref{fig1}(b)) \cite{Riek2015, Riek2017}. 
This feature indicates strong quantum fluctuations of the field on a sub-cycle timescale. As a result, while classical light can drive electrons and quasiparticles with only a well-defined quasimomentum $k$ at the time, BSV pulses can excite and drive quasiparticles in a superposition of quasi-momentum states. We illustrate the difference between the two cases in (Fig.1 (c)) and (Fig.1 (d)) by depicting BSV-excited quasiparticles moving simultaneously in opposite directions (Fig. 1 (d)).  
Accordingly, within half of the optical cycle, the electron-hole pairs can be driven to spread all over the energy bands, emitting a broad spectrum of harmonic radiation. 
Large variance of the BSV drive also implies that each pulse has a Keldysh parameter~\cite{Keldysh1965} with a large uncertainty, thus affecting the competition between multi-photon and tunneling excitation.

\section{Spectra of harmonics driven by quantum light}\label{sec2}

\begin{figure}[h]
\centering
\includegraphics[width=1.0\textwidth]{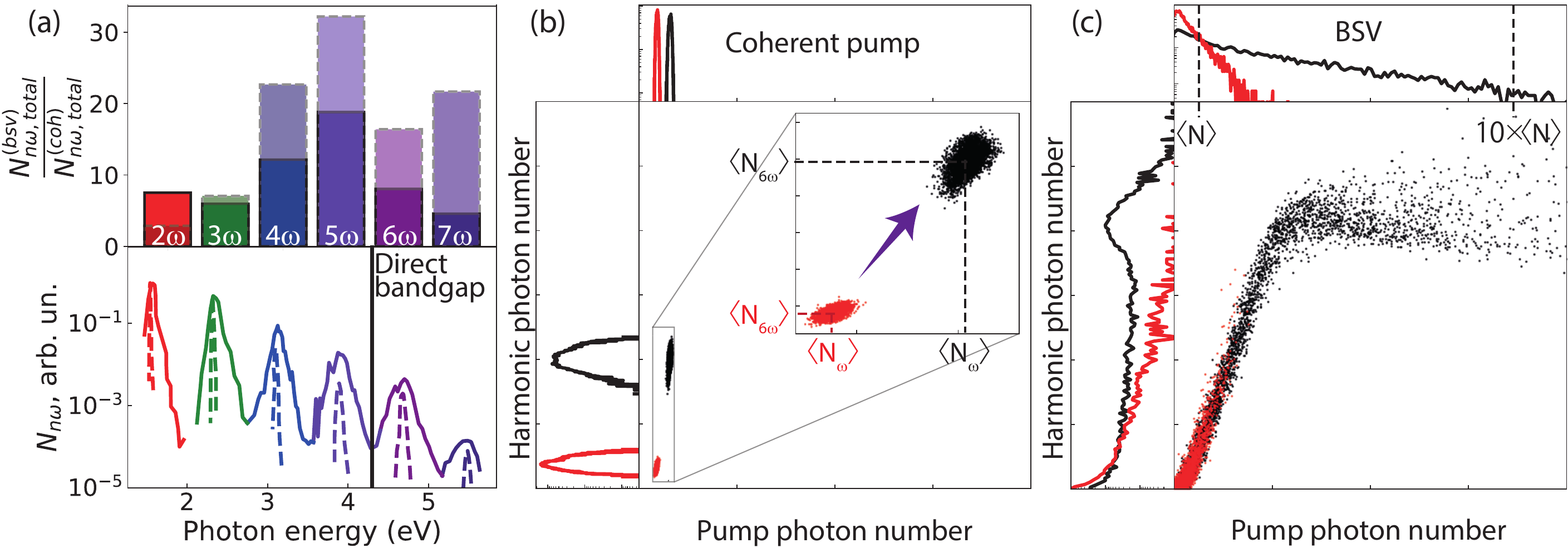}
\caption{(a) The spectrum of harmonics generated in a 6 $\mu$m x-cut doped lithium niobate crystal by bright squeezed vacuum (solid line) and by coherent radiation (dashed line). The photon energies of the 6th and the 7th harmonics exceed the direct energy bandgap of doped lithium niobate. Upper panel: Harmonic yield enhancement factor BSV vs.~coherent pump for the same mean intensity ($\sim 2$ $\mathrm{TW\,cm}^{-2}$). We show experimental (solid bars) and simulated (dashed bars) factors. (b,c) Joint photon-number probability distribution of the input pump (b, coherent radiation, c, bright squeezed vacuum) and the 6th harmonic for two different mean pump photon numbers (red and black). In the case of coherent pump, the joint photon-number distribution is Gaussian while the photon statistics in the BSV case show non-trivial features.}\label{fig2}
\end{figure}

We compare HHG from solid targets driven by ultrashort pulses of classical coherent and quantum light. Both pulses are centred at 1.6 $\mu$m (0.77 eV), with the former having a duration of 70 fs at full-width half maximum (FWHM) and the latter of 25 fs (FWHM). As targets we use: 6 $\mu$m-thick x-cut Mg:LiNbO$_3$ (LN) and 1 $\mu$m-thick amorphous silicon (a-Si), observing emission of harmonics up to the 7th order. 

The HHG spectrum of LN (Fig.~\ref{fig2}(a)) shows both odd and even harmonics, due to the non-centrosymmetric crystal structure.
The photon energies of the 2nd to the 5th harmonic from LN are below the direct bandgap (4.2 eV \cite{LN_bandgap}), and the 6th and 7th harmonics are above the direct bandgap (Fig.~\ref{fig2}(a)).
Despite the shorter duration, the BSV-generated  4th-7h harmonics show at least 5 to 15 times higher yield than the same harmonics driven by coherent pulses with identical mean intensity ($2 \hbox{ TW}\ \hbox{cm}^{-2}$).

Simulations based on the semiconductor-Bloch equations and accounting for the BSV photon statistics shows a similar behavior, an overall increase of the enhancement up to the 5th harmonic and a lower enhancement for 6th and 7th harmonics (see Fig.~\ref{fig2}(a) and Methods). 

Besides measuring the mean value of the harmonic yield, we also study the joint photon-number statistics between the input pump light and the harmonic radiation. For this purpose, we measure the shot-to-shot input light intensity and the resulting harmonic intensities (see Methods).
Figures \ref{fig2}(b,c) show the measured two-dimensional photon-number distributions between the pump and the 6th harmonic from LN in the case of the coherent pump (panel b) and BSV pump (panel c) for two different mean photon numbers (red and black points).
The photon-number probability distribution for classical light is narrow and Gaussian, centered at the mean photon number $\langle N_{\omega} \rangle$. 
The photon-number probability distribution of the harmonic is also Gaussian with the mean photon number $\langle N_{6\omega} \rangle$ (Figure \ref{fig2}(b)).
The increase of the pump's mean photon number shifts the center of the joint probability distribution without significantly changing its width.
The measurement of the mean photon number is thus sufficient to find the harmonic yield $\langle N_{6\omega} \rangle/\langle N_{\omega} \rangle$, but obtaining the power scaling requires a systematic scan of the pump mean photon number.

As opposed to coherent radiation, BSV has a broad photon-number distribution with its maximum at zero photon number.
The photon-number distribution of the harmonic is non-trivial and differs from that of the driving pulses but it inherits the large width; moreover, there is a correlation between the BSV photon number and the harmonic photon number (Fig.~\ref{fig2}(c)). 
As a result, the large variance of the BSV photon number reveals the entire power dependence through a single joint probability distribution measurement. In other words, a scan of the mean photon number of the BSV excitation is not required, underscored by the fact that both low and high BSV setting (red vs black in Figure \ref{fig2}(c)) can reveal the entire power dependence.

Additionally, within a single optical cycle of BSV, the superposition of its photon-number states is mapped to the quasi-momentum of an electron; therefore, its wavefunction is distributed over the whole energy band.
This implies that the joint photon-number distribution between the BSV and its harmonic, which results from the electron-hole recombination, contains information of a large part of the (or even the entire) sample band structure and electron dynamics.

\section{Power scaling of the high harmonics}\label{sec3}

\begin{figure}[h]
\centering
\includegraphics[width=1.0\textwidth]{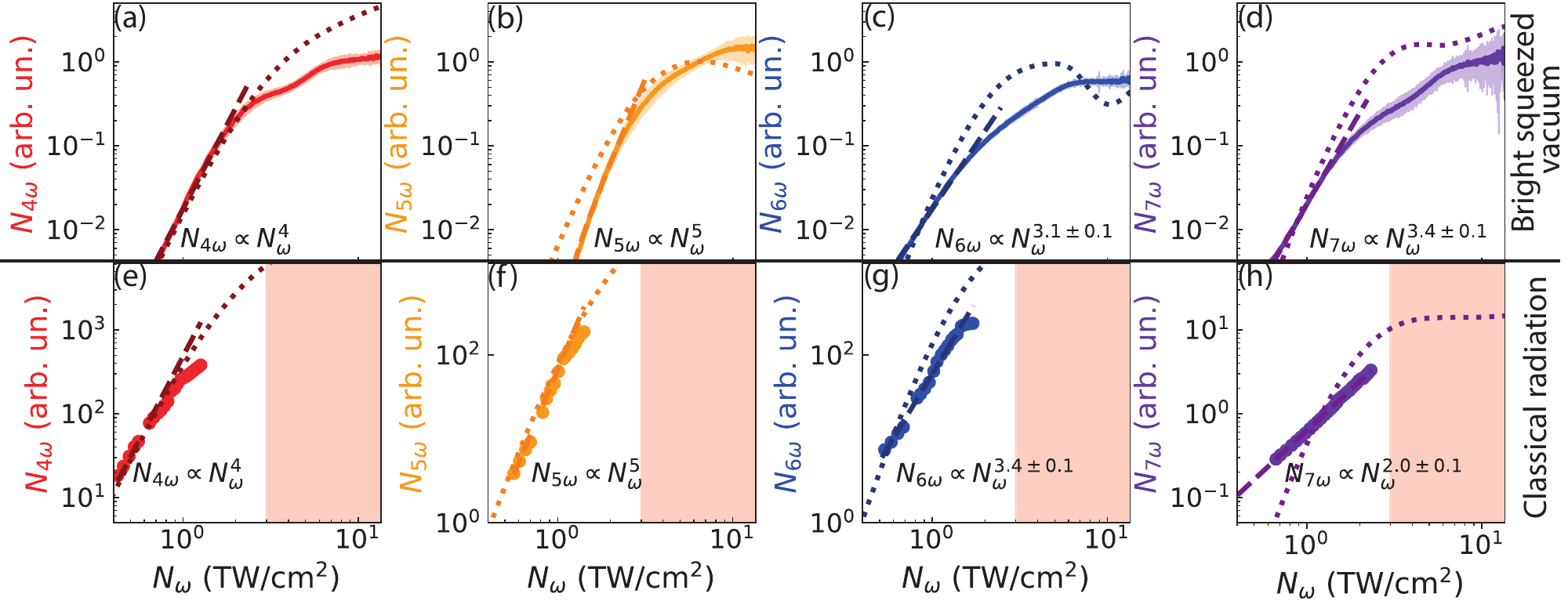}
\caption{Power scaling of the 4th-7th harmonics generated in x-cut lithium niobate by bright squeezed vacuum (a-d) and by coherent radiation (e-h). Solid lines: mean photon number of the harmonics, with the shading indicating the photon-number uncertainty. Dotted lines: numerical simulation. Dashed lines: power law fit. For coherent light excitation, the range above 3 TW cm$^{-2}$ is inaccessible because of the sample damage. }\label{fig32}
\end{figure}

In Fig.~\ref{fig32}, we show the yield of the 4th-7th harmonics generated by pumping LN with BSV light (top row) and classical light (bottom row) measured per each laser shot and the corresponding simulations. At low pump intensities, the power of harmonics with frequency below the direct bandgap (4th and 5th) scales with the power equal to the harmonic order, typical of the perturbative regime. As the intensity increases, the scaling changes and becomes non-perturbative. When using the classical pump, deviations from the perturbative scaling can be observed at intensities below 1  TW cm$^{-2}$ (Fig.~\ref{fig32}(e,f)). 

In contrast, the same harmonics generated by BSV exhibit a perturbative power scaling over a broader range of intensities and a deviation from it at around 2 TW cm$^{-2}$.
The yield of the 6th and 7th harmonics (with frequencies above the direct bandgap) follows a non-perturbative power scaling over the whole intensity range with both coherent and BSV pumps, but with some differences (Fig.~\ref{fig32}(c,d,g,h)). The classical-light generated harmonics scale with a fixed exponent, but this one changes with intensity when generating the harmonics with BSV light. While at low intensities, the 6th harmonic power scales with a comparable exponent ($\sim$3.4) for both pumps, we observe two different exponents for the 7th harmonic: 2.0 with coherent and 3.4 with BSV. 
Above 1.5 TW cm$^{-2}$, the scaling exhibits two plateaus, especially pronounced for the 4th and 7th harmonics, likely originating from quantum path  interference of short and long trajectories generated in the HHG process~\cite{Kim2019}. These features are visible only in the case of the BSV pump, due to its possibility of `non-invasive' testing. Numerical simulations can reproduce our experimental results fairly well for both classical and BSV excitations (see Fig.~\ref{fig32} and Methods).
At low pump intensities (below $\sim$ 2 TWcm$^{-2}$) and for the 4th - 6th harmonics, the agreement between experimental data and simulations is excellent. However, the two depart from each other as the intensity increases, and for the 7th harmonic in general. 
We attribute these deviations to the simplified bandstructure model, where we only take into account a single conduction band (see Methods and Supplementary Information).

\begin{figure}[h]
\centering
\includegraphics[width=0.7\textwidth]{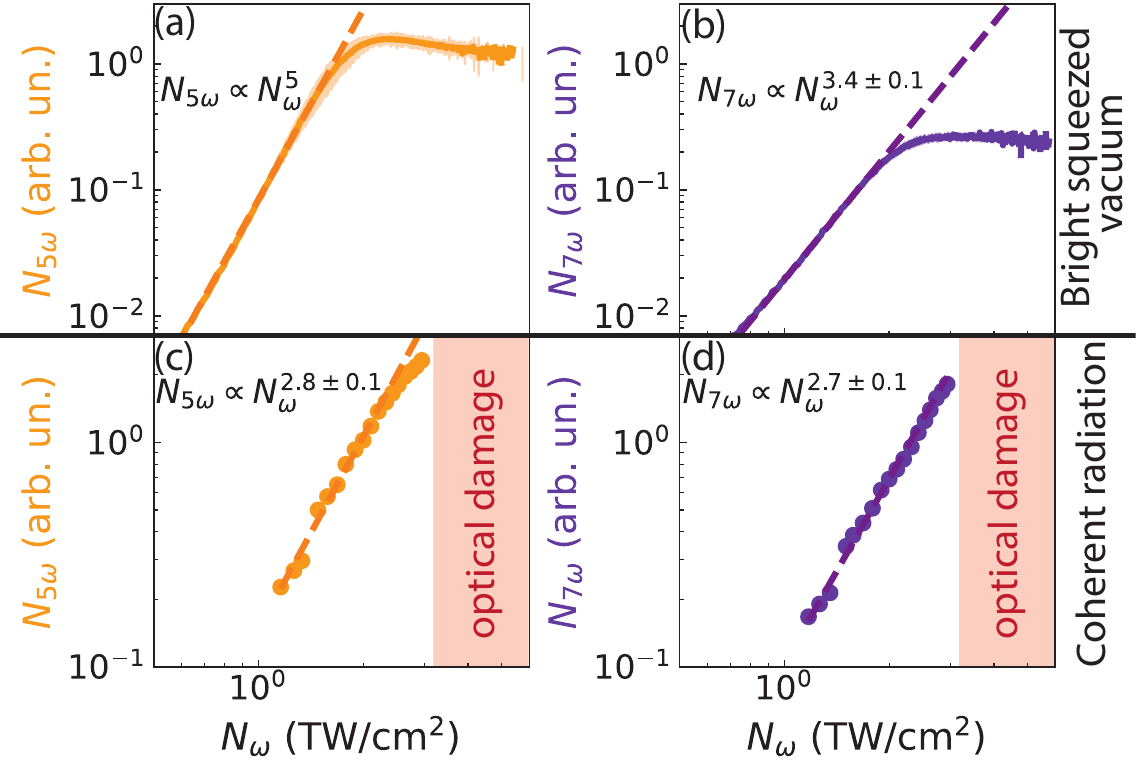}
\caption{Power scaling of the 5th and 7th harmonics generated in amorphous silicon by bright squeezed vacuum (a,b) and by coherent radiation (c,d). Solid lines: mean photon number of harmonics, with the shading showing the photon-number uncertainty. For coherent light excitation, the range above 2 TW cm$^{-2}$ is inaccessible because of the sample damage.}\label{fig31}
\end{figure}
The differences in the power scaling of harmonics driven by BSV and by classical light become even more striking when we replace LN with a-Si, which has a lower direct bandgap (in the range between 1.5 eV and 2 eV). Because of the amorphous material structure, only odd harmonics can be generated from this sample.
Figure \ref{fig31} shows the (shot-to-shot) power scaling of the above-bandgap 5th and 7th harmonics generated by BSV (a,b) and classical radiation (c,d) in a-Si.
Under coherent-state irradiation of the sample, we observe a non-perturbative  scaling of the photon number for both harmonics: $N_{5\omega} \propto N_{\omega}^{2.8}, N_{7\omega} \propto N_{\omega}^{2.7}$.
However, when using BSV, we obtain a different scaling. The yield of the 5th harmonic initially scales with the 5th power of the pump photon number, $N_{5\omega} \propto N_{\omega}^5$, indicating a perturbative scaling, and then saturates at pump intensities above $\sim 2$ TW cm$^{-2}$. 

Although the yield of the BSV-generated 7th harmonic already shows the non-perturbative scaling as a sign of tunneling scenario, its exponent is larger than in the case of coherent pump, which points to a possible competition between the 7-photon multiphoton process and the non-peturbative regime of HHG.

We attribute the differences in the power scaling of the two types of light to the superbunched photon statistics of BSV, which leads to the enhancement of $n$-photon processes by a factor of $g_{BSV}^{(n)}$, compared to coherent-state (classical) light \cite{Spasibko2017}. This enhancement offers an explanation for the broader intensity range over which the harmonic yield follows a perturbative scaling as well as the higher exponent of the BSV-generated 7th harmonic at low pump intensities.
Furthermore, this interpretation also explains the lower statistical enhancement of the harmonics generated above the direct bandgap compared to the 5th harmonic (Fig.~\ref{fig2}a). A power scaling with an exponent around 3 leads to an enhancement by only a factor of $g_{BSV}^{(3)}=15$, instead of $g_{BSV}^{(5)}=945$ in the case of the 5th harmonic.  

Another feature of HHG driven by BSV is the saturation of the yield, which we attribute to the strong depletion of the valence band population. Remarkably, this is not observed in the case of coherent-state pump due to the optical damage of the sample.

\section{Discussion}\label{sec12}

The harmonic power scalings obtained with both samples show that using BSV we generate harmonics over a wide range of pump intensities extending beyond 10 TW cm$^{-2}$ without causing optical damage. On the other hand, with coherent light, considerably lower pump intensities ($\sim$ 2 TW cm$^{-2}$) can induce damage in the samples. Such striking difference cannot be entirely explained by the longer duration of the coherent pulses and is mostly a consequence of the diverse photon distribution of the two pulses. 

Under coherent light irradiation, each pulse contains a well-defined amount of photons, which will cause optical damage to the sample above a certain threshold. In contrast, BSV pulses possess a broad photon-number distribution, implying that each pulse contains an uncertain number of photons, with the mean value far below the damage threshold. The large-photon-number events in the tail of the distribution, however, can drive HHG and probe the material without damaging it within a single laser shot or multiple laser shots. Often, continuous irradiation of the sample within temporal intervals shorter than its relaxation time can result in optical damage to the sample. The BSV photon-number distribution also implies a longer temporal interval between intense events, thus effectively providing the sample with sufficient time to relax to its initial state.

Additional experiments (not shown) reveal that by decreasing the laser repetition rate, the sample damage threshold increases, and we can generate high harmonics using brighter classical light. However, covering a comparable intensity range as BSV requires reducing the repetition rate to the Hz level, increasing significantly the already longer acquisition time. The observation indicates that opportunely modulating the shot-to-shot power of a coherent source would allow using it for HHG and acquiring data within times and intensity ranges comparable to those achieved with BSV. However, this would come at the cost of increasing complexity, especially for sources of few-cycle pulses operating at high repetition rates. On the other hand, the natural sparsity of BSV provides an alternative and more powerful route to using optical modulators to study the interaction of intense light with matter, allowing, in a fashion similar to sparse sampling, the collection of sufficient data within a short time (minutes using a 1 kHz repetition rate pump laser) without damaging the solid sample nor adding any experimental complexity.

Thus, the excitation of a material with BSV is more gentle than with coherent light. It enables the study of solid-state samples in extreme regimes, such as the harmonic yield saturation in amorphous silicon (Fig.~\ref{fig31}(a, b)), which can be related to the creation of an electron-hole plasma.

\section{Conclusion and outlook} 

In conclusion, our observation of non-perturbative HHG driven by BSV demonstrates that experiments with quantum light are now able to access the strong-field regime of quantum optics. BSV enables a strong enhancement of solid-state HHG compared to coherent light and allows for the observation of electron-hole dynamics beyond the conventional damage threshold. The approach we developed may become a new spectroscopic tool based on photon-number statistics, heralding the emergence of extreme nonlinear quantum spectroscopy. Our BSV source can also attain sufficient intensities to drive HHG in gases, liquids, and wide-bandgap solids, reaching the XUV spectral regime, and we expect that can also be employed for controlling HHG, based on sub-cycle quantum noise engineering, in a similar fashion to control through sub-cycle field engineering.

In the future, the giant uncertainty of the BSV's electric field magnitude and the massive amount of photon pairs contained in each BSV pulse will find use in the observation of quantum interference effects and of many-body correlations in solids, with applications in quantum state engineering.
Finally, since HHG occurs on sub-cycle timescales, it goes beyond conventional descriptions of single-mode quantum optics, which is insufficient over such short temporal spans. Our findings will further stimulate discussions in this direction and expand the realm of quantum and extreme nonlinear optics to new uncharted territories.

\section{Methods}\label{sec11}

\subsection{Experimental setup}

\begin{figure}[h]
\centering
\includegraphics[width=0.8\textwidth]{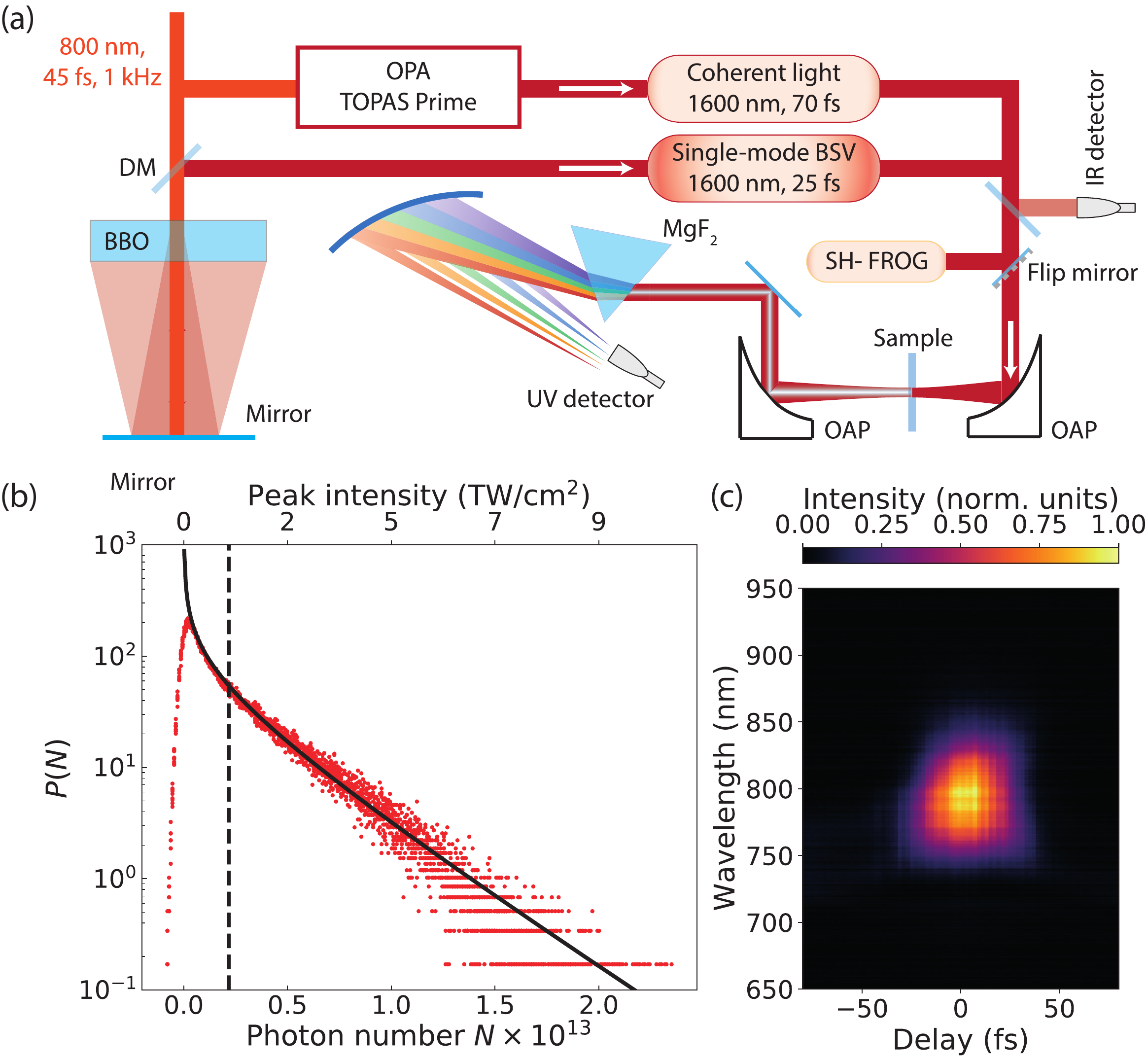}
\caption{(a) Experimental setup for HHG by coherent radiation and BSV. (b) Photon-number distribution of BSV. The solid line is a theoretical fit, which corresponds to single-mode BSV, the only fitting parameter being the mean photon number (dashed line). (c) Measured averaged FROG trace of BSV.}\label{fig_setup}
\end{figure}

Figure \ref{fig_setup}(a) shows a sketch of our experimental setup for HHG in solids.
For comparison, we use two linearly polarised pump sources delivering (i) coherent pulses and (ii) BSV pulses both centered at 1600 nm. As a pump laser for both sources, we use a titanium sapphire amplifier laser system (Coherent Legend) with a central wavelength of 800 nm,  45 fs  FWHM pulse duration, and a repetition rate of 1 kHz. Combining this laser with a femtosecond optical parametric amplifier (Light Conversion, TOPAS Prime), we obtain coherent pulses with a duration of $70$ fs (FWHM).

In a parallel setup, after reducing the pump laser beam diameter to 3 mm (1/$e^2$), we generate BSV in a $3$-mm-thick  BBO crystal  cut for type-I collinear frequency-degenerate phase matching via high-gain parametric down-conversion. Here, the BBO crystal is tuned out of the exact phase matching to reduce the frequency bandwidth of the BSV, so that the BSV becomes single-mode temporally.
However, at this stage, BSV pulses still contain around 150 spatial modes.
In order to make it single-mode also spatially, we reflect both the BSV and the pump back into the same crystal with a plane silver mirror placed at a distance of $75$ cm from the crystal. Back at the BBO, only the BSV spatial mode with the lowest diffraction (e.g., the fundamental mode) overlaps significantly with the pump beam and undergoes phase-sensitive amplification, while all the higher-order modes diffract faster without being further amplified~\cite{Perez:14}.
After the amplification, the BSV pulses exhibit a spectral bandwidth of 150 nm centered around 1.6 $\mu$m. We separate them from the residual 800 nm light using a shortpass dichroic mirror (Thorlabs DMSP1180) and two longpass filters (Thorlabs FELH1000).

To characterize the temporal profile of the coherent and BSV pulses, and thus determine the actual intensity of the light irradiating the sample, we use a home-build device for second-harmonic frequency-resolved optical gating (SH-FROG). The retrieved FROG traces reveal that both pulses are almost transform-limited with a duration of 70 fs FWHM for the coherent pulses and 25 fs FWHM for the BSV pulses. The measured FROG trace of the latter is shown in  Fig.~\ref{fig_setup}(c).

Figure~\ref{fig_setup}(b) displays the photon-number distribution of BSV measured with a fast InGaAs photodiode (Thorlabs PDA20C2) without spatial or spectral filtering.
The data are well fitted with a theoretical single-mode distribution \cite{Spasibko2017} (black line), with the mean number of photons per pulse $2\cdot10^{12}$ (dashed vertical line), confirming that the BSV is single-mode. 
The distribution stretches up to photon numbers of $2\cdot10^{13}$, corresponding to pulse energies of 9 $\mu$J. 

For HHG, either the BSV or the coherent-state radiation is focused into a 30 $\mu$m spot (diameter at $1/e^2$) on the sample with a silver off-axis parabolic mirror (Edmund Optics, $f=25.4$ mm) in order to drive HHG. The spot size has been determined from a knife-edge measurement for both sources.  
We use two samples: an x-cut  6 $\mu$m Mg:LiNbO$_3$ crystal on a $1$ mm fused silica substrate and a 1 $\mu$m a-Si layer on a $1$ mm fused silica substrate. We mount the samples so that the substrate faces the pump pulses. Because fused silica has a significantly higher bandgap, it does not contribute to the outcome of the HHG measurement.
After the samples, the harmonics and the pump light are collimated by a UV-enhanced aluminum off-axis parabolic mirror (Thorlabs MPD119-F01, $f=25.4$ mm) and sent to a MgF$_2$ prism. After this, the harmonics are focused by a UV-enhanced aluminum spherical mirror (Thorlabs CM254-250-F01) to different positions.

We detect the 4th--7th harmonics by a fast UV-enhanced silicon avalanche photodiode (Thorlabs APD440A2) separately by rotating the spherical mirror within a range of 1°.
Additionally, we place bandpass filters for the 4th harmonic (Thorlabs FBH400-40), the 5th harmonic (Edmund Optics 12094, BP325/50) and the 6th harmonic (Edmund Optics 39313, BP266/10), and also an iris (1 mm opening) in front of the detector to reduce the contribution of the scattered light.
We acquire 200,000 single-shot records of driver pulses (classical or BSV) and each harmonic synchronously.

\subsection{Numerical simulation}

To simulate and elucidate our experimental findings, we utilize the semiconductor Bloch equations (SBEs)~\cite{luu2016high} to describe the interaction between BSV (or classical) light and LiNbO$_3$ along $\Gamma$-$Z$:

\begin{equation}
\frac{\partial P(k,t)}{\partial t} = [\epsilon_c(k) - \epsilon_v(k) + iE(t) \nabla_k]P(k,t) - (1 - f_k^e - f_v^h)d_k \cdot E(t) - i \frac{1}{T_2} P(k,t),
\end{equation}

\begin{equation}
\frac{\partial f_k^{e(h)}(k,t)}{\partial t} = -2\text{Im}[d_k \cdot E(t) \cdot P_k^*] + eE(t) \nabla_k f_k^{e(h)}(k,t).
\end{equation}

Here, $P(k,t)$ represents a dimensionless polarization depending on the time $t$ and the quasi-momentum $k$. $f_k^{e(h)}$ denotes the population of electrons and holes in the conduction and valence bands, respectively. The energy-momentum relation for the conduction band $\epsilon_c(k)$ and the valence band $\epsilon_v(k)$, as well as the transition dipole moment (TDM) $d_{k}$, are derived from density functional theory (DFT) calculations~\cite{shao2022spontaneous}, with some modifications. Most importantly, we set the bandgap manually to 4.2 eV, corresponding to the direct bandgap of doped LN, which is the relevant gap for the HHG process. The exact bandstructure and transition dipole moments are plotted in the Supplementary Information.

To eliminate multiple recollision interferences and maintain the pronounced harmonic structure in our 1D simulation, we set the phenomenological decoherence time to half of the optical cycle duration of the pump laser, \(T_2 = 0.5T_{\text{pump}}\). Extending the decoherence time to one optical cycle \(T_{\text{pump}}\) does not notably alter our results. In general, a longer decoherence time exceeding a few cycles allows electrons to return over longer durations than a laser cycle, leading to unphysical anharmonic photon emission, whereas a shorter decoherence time less than a quarter cycle \(0.25T_{\text{pump}}\) may significantly suppress long trajectories.

Considering LiNbO$_3$'s ferroelectric nature, we account for the spontaneous polarization effect, which breaks the material's symmetry and allows for even-order harmonics generation (see Ref.~\cite{shao2022spontaneous} and the Supplementary Information for more details). The HHG spectrum is computed as 
\begin{equation}
S_{\text{HHG}}(\omega) \sim \left| \int_{-\infty}^{\infty} [J_{\text{inter}} + iJ_{\text{intra}}] e^{i\omega t} dt \right|^2,
\end{equation}
where 
\begin{equation}
J_{\text{inter}} = \frac{d}{dt} \int d_k \cdot P(k,t) dk + \text{c.c.}
\end{equation}
and
\begin{equation}
J_{\text{intra}} = \sum_{n=c,v} \int \nu_n(k) \cdot f_n(k,t) dk
\end{equation}
denote, respectively, the interband and intraband contributions to the HHG yield,  and $\nu_n(k) = \nabla_k \epsilon_n(k)$ represents the group velocity of the electrons.

For simulating the BSV-driven HHG spectrum underlying the enhancement plot (Fig.~\ref{fig2}(a)), we employ the Husimi Q function for BSV~\cite{Gorlach2023}:
\begin{equation} \label{Q_distribution}
Q(\epsilon_\alpha) \approx \frac{2}{\sqrt{2\pi|\bar{\epsilon}|^2}} \exp\left(-\frac{|\epsilon_\alpha|^2}{2|\bar{\epsilon}|^2}\right).
\end{equation}
Here, $\epsilon_\alpha$ is the electric field amplitude for coherent parameter $\alpha$, and $\bar{\epsilon}$ the mean value of the absolute field amplitude of BSV. $\epsilon_\alpha$ is taken to be real along the direction of antisqueezing, and the distribution is integrated over the axis of squeezing in advance, so that the distribution in Eq.~\ref{Q_distribution} is one-dimensional. This is justified as the squeezed axis of the Husimi distribution is infinitely narrow in terms of the electric field amplitude, and hence does not contribute to the spectrum.

The BSV-driven HHG spectrum is obtained by integrating the coherent HHG spectra with the Q function distribution~\cite{Gorlach2023}:

\begin{equation}
S^{\text{BSV}}_{\text{HHG}}(\omega,\bar{\epsilon}) =\int d\epsilon_\alpha Q(\epsilon_\alpha)S^{\text{coh}}_{\text{HHG}}(\omega,\epsilon_\alpha).
\end{equation}

\backmatter

\bmhead{Data availability}
All data that support the plots within this paper and other findings of this study are available from the corresponding author upon reasonable request.

\bmhead{Acknowledgements}
F.T., M.C.~and A.R. thank Philip St.J. Russell for supporting the project.
The authors thank Matan Even Tzur and Misha Ivanov for fruitful discussions.
A.R. thanks Patrick Cusson for the help with the FROG measurement and Isaac Soward for the help at the first stage of the experiment.
A.R. acknowledges funding from the International Max Planck Research School for Physics of Light. 
D.S. acknowledges partial support by the European Union's Horizon Europe Research and Innovation Programme under agreement 101070700 (project MIRAQLS).
Z.C., M.B., O.C., I.K.~and M.K.~thank the Helen Diller Quantum Center for partial financial support.

\bmhead{Author contributions}
M.C.~and F.T.~conceived the project, supervised the work and acquired the funding. A.R., F.T.~and M.C.~designed the experiment. A.R. carried out the experiment under supervision of M.C., F.T. and D.S. Z.C.~and M.B.~developed the theory and carried out the numerical simulations under the supervision of M.K., I.K.~and O.C. A.R., M.C., D.S., M.K.~and F.T.~wrote the article. All authors contributed to discussions and the interpretation of the results.

\bmhead{Declarations}
The authors declare no competing interests.

\bibliography{HHG_literature}

\end{document}